\documentclass[final,3p,times,twocolumn]{elsarticle}
\usepackage{blindtext, graphicx, amsmath, algorithm, algpseudocode, pifont, algcompatible, layout, amsthm, amssymb}
\usepackage{mathrsfs}
\usepackage{mathtools}
\usepackage{enumitem}   
\usepackage[dvipsnames]{xcolor}
\usepackage{eso-pic}
\usepackage{booktabs}
\usepackage{color}
\usepackage{graphicx}
\usepackage{float}
\usepackage{caption}
\usepackage{subcaption}
\usepackage[ulem=normalem]{changes}
\usepackage{tikz,pgfplots}
\usepackage{url}
\pgfplotsset{compat=newest}
\usetikzlibrary{graphs,matrix,positioning,calc,fit,backgrounds,chains}
\usepackage[breakable]{tcolorbox}

\usepackage[utf8]{inputenc}
\usepackage[english]{babel}
\usepackage{hyperref} 
\hypersetup{ colorlinks=true, linkcolor=black, filecolor=black, urlcolor=cyan, }

\usepackage{caption}
\let\oldbrace\{
\def\{{\oldbrace\kern0.5pt}

\newcommand{\Cc}{\mathcal{C}}

\newcommand{\Lc}{\mathcal{L}}
\newcommand{\Mc}{\mathcal{M}}
\newcommand{\Nc}{\mathcal{N}}

\newcommand{\Xc}{\mathcal{X}}
\newcommand{\Yc}{\mathcal{Y}}

\DeclareMathOperator\E{\sf E}

\newcommand{\Complex}{\mathbb{C}}

\newcommand{\Field}{\mathbb{F}}

\theoremstyle{definition}

\newcommand{\SNR}{\mathsf{SNR}}

\newcommand{\ms}{\mathsf{m}}

\journal{ICT Express}

\begin{document}

\begin{frontmatter}

\title{Hybrid Neural Coded Modulation: Design and Training Methods}
%\author{Author’s Full Name 1\corref{cor1}, Author’s Full Name 2, Author’s Full Name 3}
\author[1]{Sung Hoon Lim}
\ead{shlim@hallym.ac.kr}

\author[1]{Jiyong Han}
\ead{jyong0719@gmail.com}

\author[1]{Wonjong Noh}
\ead{wonjong.noh@hallym.ac.kr}

\author[2]{Yujae Song}
\ead{yjsong@kiost.ac.kr}

\author[3]{Sang-Woon Jeon\corref{cor1}}
\ead{sangwoonjeon@hanyang.ac.kr}

\address[1]{School of Software, Hallym University, Chuncheon, Korea}
\address[2]{Maritime ICT R\&D Center, KIOST, Busan, Korea}
\address[3]{Department of Electrical and Electronic Engineering, Hanyang University, Ansan, Korea}

\cortext[cor1]{Corresponding author}
%\ead{author1@ictexpress.com, author2@ictexpress.com, author2@ictexpress.com}

\allowdisplaybreaks

\begin{abstract}
We propose a hybrid coded modulation scheme which composes of inner and outer codes. The outer-code can be any standard binary linear code with efficient soft decoding capability (e.g. low-density parity-check (LDPC) codes). The inner code is designed using a deep neural network (DNN) which takes the channel coded bits and outputs modulated symbols. For training the DNN, we propose to use a loss function that is inspired by the generalized mutual information. The resulting constellations are shown to outperform the conventional quadrature amplitude modulation (QAM) based coding scheme for modulation order 16 and 64 with 5G standard LDPC codes.
\end{abstract}


\end{frontmatter}

%%\textbf{}
% \linenumbers

\section{Introduction}\label{sec: introduction}
Machine learning methods for channel coding is an emerging field which can help overcome many challenging problems in error-correction coding. In particular, end-to-end learning for designing encoders and decoders have been proposed in~\cite{oshea--hoydis2017} that utilizes a deep neural network (DNN) autoencoder. An interesting insight of this approach is that the end-to-end structure does not utilize conventional quadrature amplitude modulation (QAM) constellations nor does it make use of the de-facto architecture, bit interleaved coded modulation (BICM)~\cite{caire--taricco--biglieri1998, Martinez--Guillen--Caire--Willems2009}. Instead, it is a clean slate approach to find the optimal encoder and decoder pair using a DNN. One benefit of this approach is that the encoder is not restricted to the suboptimal QAM constellation and may learn a better input distribution overall. End-to-end transceiver design utilizing DNNs has been applied in various contexts including additive white Gaussian noise (AWGN) channels~\cite{ Dorner--Cammerer--Hoydis--Brink2018, Jiang--Kim--Asnani--Kannan--Oh--Vishwanath2019, He--Jin--Feifei--Li--Xu2019}, fast fading~\cite{Park--Simeone--Kang2019}, intersymbol interference (ISI) channels~\cite{Zhang--Wu--Coates2021}, ultra low-latency~\cite{jiang--kim--asnani--kannan--oh--viswanath2020}, and model free design~\cite{Aoudia--Hoydis2019}. Other works have also used DNNs focusing on specific components such as decoder design~\cite{Nachmani--Marciano--Burshtein2018, Shental2019,Koike-Akino--Wang--Millar--Kojima--Parsons2019,  He--Jiang--Lin--Zhao2020, Shah--Vasavada2021, Dai--Tan--Si--Niu--Chen--Poor2021, Nachmani--Wolf2021} and constellation shaping for modulation~\cite{Stark--Aoudia--Hoydis2019}. Several other approaches to canonical problems have been proposed for feedback channels~\cite{kim2020}, quantized channel observations~\cite{Balevi--Andrews2020}, joint source--channel coding~\cite{Saidutta2021}, and the wiretap channel~\cite{Besser2020}.
More recently, theoretical studies on end-to-end design have been given in~\cite{Weinberger2021}. While these approaches provide breakthroughs in various situations, one challenge in the end-to-end design approach is that the exponential growth on the number of codewords in code length makes learning end-to-end codes quite difficult for long codes.

In another line of work, hybrid architectures were proposed which consist of a DNN inner code (or modulator) that is concatenated to an outer linear code (e.g. turbo and low-density parity-check (LDPC) codes). For example,
hybrid architectures were designed and applied for AWGN with radar interference~\cite{Alberge2019}, optical fiber communications~\cite{jones--Yankov--Zibar2019}, one-bit quantized AWGN channels~\cite{Balevi--Andrews2020}, and AWGN channels~\cite{Cammerer--Aoudia--Dorner--Stark--Hoydis--Brink2020}. The work of~\cite{Cammerer--Aoudia--Dorner--Stark--Hoydis--Brink2020} also generalizes the decoder for iterative demodulation and decoding (IDD) and gives implementation results on software defined radios. In these approaches, the binary cross entropy (BCE)~\cite{Murphy2012} metric was used to train the DNN which enables the inner code to be compatible with the outer linear code decoder, i.e., the DNN output is in the form of bit-level decoding metrics.
An advantage of the hybrid structure approach is that it can benefit from learning a better constellation (shaping gain) while maintaining practical code lengths for error correction performance (coding gain).

Following this approach, we propose a generic architecture that can take some off-the-shelf linear code (e.g. LDPC codes) and concatenate it with a DNN autoencoder inner-code for modulation. With this goal in mind, we design and train a DNN inner-code that is compatible for the linear channel code decoder, for example, to be compatible with the sum-product algorithm (SPA). Our main contribution in this direction is that we propose an approximated formulation of the generalized mutual information (GMI)~\cite{Merhav1994, Martinez--Guillen--Caire--Willems2009} as a loss function that is tailored for learning the inner DNN encoder and decoder pair. We further outline some useful training techniques that we have learned through extensive evaluations.

In the numerical evaluations section, we provide our performance evaluations with our trained DNN inner code concatenated with a 5G standard LDPC code~\cite{3gpp} and compare its performance with QAM based BICM systems for modulation order $M=16$ and $M=64$.

%Some related works in this direction are~\cite{jones--Yankov--Zibar2019} which utilizes the binary cross entropy function as an upper bound for GMI maximization on optical fiber communications,~\cite{Balevi--Andrews2020} uses a hybrid structure for one-bit quantized AWGN channels, and~\cite{Shental2019} trains a DNN for soft demodulation.\textbf{}

In the sequel, we define $\Complex$ and $\Field_2$ as the complex and binary field. Random variables are denoted by upper-case letters $X$, and expected values are noted by $\E[X]$. We define $c^n := (c_1,\ldots, c_n)$, i.e., an $n$-length sequence (or vector) and will often re-index a sequence $c^n$ by its $j$-th subsequence $c^{(j)} = (c^{(j)}_1,\ldots, c^{(j)}_m)$ of length $m$ such that $c^n = (c^{(1)}, \ldots, c^{(N)})$, where $N = n/m$. The length of the subsequence will be noted in the context when needed.

\section{System model}\label{sec: system}

\subsection{Communication system and channel model} \label{sec: channel model}
Consider a memoryless channel 
$(\Xc, p_{Y|X}, \Yc)$
which consists of an input alphabet $\Xc$, a receiver alphabet $\Yc$, and a collection of conditional distributions $p_{Y|X}$. 

A $(2^{NR},N)$ code for the channel consists of a message set $[1:2^{NR}]$, an encoder which maps each message $\mathsf{m} \in [1:2^{NR}]$ to a sequence $x^N(\ms)\in\Xc^N$, and a decoder that assigns estimates $\hat\ms(y^N) \in \Mc$ to each received sequence $y^N \in \Yc^N$. From the memoryless channel assumption we have
$P(y^N|x^N) = \prod_{i=1}^N p_{Y|X}(y_i|x_i)$ where
$p_{Y|X}$ is the AWGN channel, i.e., %$y_i = x_i + Z_i, \quad i\in[1:n]$.
\begin{align}
y_i = x_i + Z_i, \quad i\in[1:n] \label{eq:awgn}
\end{align}
where $Z_i\sim \Cc\Nc(0, \sigma^2)$ and $\Yc=\Xc=\Complex$.
We let $\sigma^2= 1/\SNR$ where $\SNR$ is the signal to noise power ratio and we assume that the input is subject to an average power constraint
$\frac{1}{n}\sum_{i=1}^n|x_i|^2\le 1$. Note that the underlined channel distribution of the AWGN channel is thus given by, 
\begin{align*}
    p_{Y|X}(y|x; \sigma^2) = \frac{1}{\pi\sigma^2}\exp\left({\frac{-|y-x|^2}{\sigma^2}}\right).
\end{align*}
%\begin{align}
%    p_{Y|X}(y|x; \sigma^2) = \frac{1}{\pi\sigma^2}e^{\frac{-|y-x|^2}{\sigma^2}}. \label{eq:awgn_pdf}
%\end{align}

% The capacity of the AWGN channel is given by,
% \begin{align*}
%     \mathsf{C} = \log(1+\SNR) \quad \text{bits/s/Hz}
% \end{align*}
% which is proved by a random coding argument using i.i.d. Gaussian sampled codes.

\subsection{Bit interleaved coded modulation} \label{sec: cm}

A coded modulation architecture specializes the generic communication system as follows. 
The messages are considered as bit sequences $b^k$, e.g., binary expansions of $\ms\in\Mc$. The encoder is specialized into two sub-components, a binary code and a bit-to-symbol mapper (modulator), and the decoder is specialized into two sub-components, symbol-to-bit level log-likelihood ratio (LLR) mapper (demodulator) and a binary channel code decoder. The input alphabet, i.e., the constellation set $\Xc$ is fixed as a discrete subset of $\Complex$ of size $M$, where $M$ is the modulation order and $m=\log(M)$.

Specifically, we denote a length $n=mN$ binary code codebook by $\Cc$ which maps $k$ binary inputs $b^k\in \Field^k_2$ to $n$ binary sequences $c^n(b^k)\in\Field^n_2$. The rate of the binary code is $R_{\Cc} = \frac{k}{n}$. The modulator then maps the $j$-th $m$-length subsequence $c^{(j)}$ of $c^n$ to a point $x^{(j)}\in \Xc$ by the function 
\begin{align*}
%\mu: \{0, 1\}^m \to \Xc
x^{(j)} = \mu(c^{(j)}), \, j=1,\ldots N.
\end{align*}
The theoretical performance of a coded modulation strategy can be measured by the \emph{generalized mutual information (GMI)}\footnote{This is a lower bound to the GMI given in \cite{Martinez--Guillen--Caire--Willems2009} with $s=1$.} in the form of 
\begin{align}
    I^\text{gmi}(X;Y) &= \E\left[\log\frac{q(X, Y)}{\sum_{x'\in \Xc }P_X(x')q(x', Y)}\right], \label{eq: gmi}
\end{align}
where $q(x, y)$ is a (symbol-level) decision metric. Note that when $q(x, y) = p_{Y|X}(y|x)$, the GMI is equal to the coded modulation capacity~\cite{Martinez--Guillen--Caire--Willems2009}. In this work, we are particularly interested in a bit-level decision metric based coded modulation strategy to be compatible with bit-level decoders (e.g. sum-product algorithm). 
To this end, we define a generic bit-level metric based demodulator that maps the $j$-th received  signal $y^{(j)}\in\Yc$ to a set of $m$ bit probabilities by the function
\begin{align*}
    \phi(y^{(j)}) = p^{(j)},\, j=1,\ldots, N
\end{align*}
where $p^{(j)}=(p^{(j)}_1, \ldots,p^{(j)}_m)$, $p^{(j)}_i \in [0,1]$ is the demodulated bit probability estimate of $c^{(j)}_i$. 

In the following section, we give a detailed description of our proposed architecture and explain how we specialize~\eqref{eq: gmi} for bit-level decoding metrics.

%Following the notation in~\cite{}, we denote the inverse mapping function for labeling position $j$ as
% $b_j : \Xc \to \{0, 1\}$, that is, $b_j(x)$ is the $j$-th bit of symbol $x$. Accordingly,
% we now define the sets
% \begin{align*}
%     X_b^j := \{x\in\Xc: b_j(x) = b\}
% \end{align*}
% as the set of signal constellation points $x$ whose binary label has value
% $b \in \{0, 1\}$ in its $j$-th position.
% More generally, we define the sets
% \begin{align*}
% \Xc_{b_{i_1}, \ldots, b_{i_v}}^{j_{i_1}, \ldots, j_{i_v}}
% \end{align*}
% as the sets of constellation points having the $v$ binary labels
% \begin{align*}
% \Xc_{b_{i_1}, \ldots, b_{i_v}}^{j_{i_1}, \ldots, j_{i_v}} := \{x\in \Xc: b_{j_{i_1}} = b_{i_1}, \ldots, b_{j_{i_v}} = b_{i_v}\}.
% \end{align*}

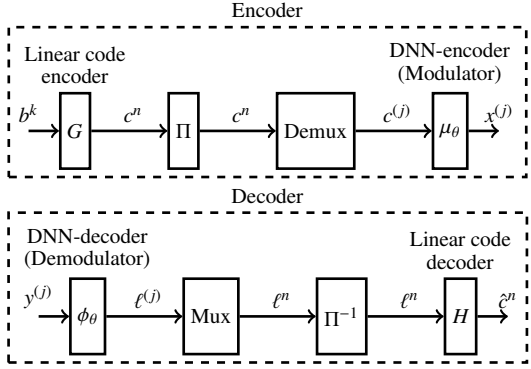
\begin{figure}[ht!]
\footnotesize
%\begin{subfigure}[b]{0.3\textwidth}
\centering
\begin{tikzpicture}[start chain, line width=1pt]
    \node[on chain, draw, minimum height=10mm,label={[name=encoder1_label,font=\footnotesize,align=center]above:Linear code\\[-\baselineskip+3mm]encoder}] (linear-enc)  {$G$};
    \node[on chain, draw, minimum height=10mm] (interleaver)  {$\Pi$};
    \node[on chain, draw, minimum height=10mm] (demux)  {Demux};
    \node[on chain, draw, minimum height=10mm,label={[name=mapping1_label,font=\footnotesize,align=center]above:DNN-encoder\\[-\baselineskip+3mm](Modulator)}] (mod 1)  {$\mu_\theta$};
    \draw[<-] (linear-enc.west) --+ (-4mm,0) node[above, label={[name=first_arrow1]}] {$b^k$};
 	\draw[->] (linear-enc.east) --+ (1mm,0) --node[above] {$c^n$} (interleaver.west);
 	\draw[->] (interleaver.east) --+ (1mm,0) -- node[above] {$c^n$} (demux.west);
 	\draw[->] (demux.east) --+ (1mm,0) -- node[above] {$c^{(j)}$} (mod 1.west);
 	\draw[->] (mod 1.east) --+ (4mm,0) node[above, label={[name=last_arrow1]}] {$x^{(j)}$};

 	\node[fit=(first_arrow1)(linear-enc)(encoder1_label)(mapping1_label)(mod 1), label={[font=\footnotesize]above:Encoder}, draw, dashed] (enc1) {};
\end{tikzpicture}
%\end{center}
%\caption{Encoder}
%\label{fig:encoder}
%\end{subfigure}
%\begin{subfigure}[b]{0.3\textwidth}
%\begin{center}
\begin{tikzpicture}[start chain, line width=1pt]
    
    \node[on chain, draw, minimum height=10mm,label={[name=demapping1_label,font=\footnotesize,align=center]above:DNN-decoder\\[-\baselineskip+3mm](Demodulator)}] (demod)  {$\phi_\theta$};
    
    \node[on chain, draw, minimum height=10mm,label={[name=encoder1_label]}] (mux)  {Mux};
    
    \node[on chain, draw, minimum height=10mm,label={[name=encoder1_label]}] (deinterleaver)  {$\Pi^{-1}$};

    \node[on chain, draw, minimum height=10mm,label={[name=decoder1_label,font=\footnotesize,align=center]above:Linear code\\[-\baselineskip+3mm]decoder}] (linear-dec)  {$H$};
	
    \draw[<-] (demod.west) --+ (-4mm,0) node[above] {$y^{(j)}$};
 	\draw[->] (demod.east) --+ (1mm,0) -- node[above] {$\ell^{(j)}$} (mux.west);
 	\draw[->] (mux.east) --+ (1mm,0) -- node[above] {$\ell^n$} (deinterleaver.west);
 	\draw[->] (deinterleaver.east) --+ (1mm,0) -- node[above] {$\ell^n$} (linear-dec.west);
 	\draw[->] (linear-dec.east) --+ (4mm,0) node[above, label={[name=last_arrow]}] {$\hat c^n$};
 	\node[fit=(demapping1_label)(linear-dec)(decoder1_label)(demod)(last_arrow), label={[font=\footnotesize]above:Decoder}, draw, dashed] (dec1) {};
\end{tikzpicture}
%\end{center}
%\caption{Decoder}
%\label{fig:decoder}
%\end{subfigure}
\caption{Proposed coded modulation system architecture. At the encoder, binary information bits $b^k$ are encoded into a binary linear code $c^n$, interleaved, and demuxed into subsequences of length $m$ $c^{(j)} = \{c^{(j)}_1,\ldots, c^{(j)}_m\}$, which is encoded by the DNN-encoder to send the $j$-th input symbol $x^{(j)}$. The decoding procedure is done in the reverse order, where $\ell^{(j)}$ are the demodulated LLRs~\eqref{eq:llr} of the bits mapped to the $j$-th symbol $x^{(j)}$, and $\ell^n$ is the multiplexed $n$ length LLRs. In the decoder, the feature mapping function $\psi$~\eqref{eq:psi} before $\phi_\theta$ and the $p^{(j)}$ to LLR $\ell^{(j)}$ transition after $\phi_\theta$ are omitted for visual clarity.}
\label{fig:system}
\end{figure}

\vspace{-0.3em}
\section{Neural network} \label{sec: neural network}

In this section, we give a detailed description of our proposed architecture for the DNN components $\mu_\theta$ and $\phi_\theta$.

A description of the proposed coded modulation system is given in Fig.~\ref{fig:system}. Note that in the figure, the modulator and demodulator are DNNs specified by parameters $\theta$. 
Our goal is to find a modulator and demodulator pair ($\mu_\theta$, $\phi_\theta$) utilizing a DNN architecture that finds the parameters $\theta$ to maximize the GMI. Once the pair ($\mu_\theta$, $\phi_\theta$) is fully trained, we treat it as an inner-code that is combined with a binary linear code (e.g. LDPC) as the outer code.

\subsection{DNN inner-encoder (modulator)}
The neural encoder $\mu_\theta$ comprises of several layers. The first input layer is a $\tanh$ layer, i.e., a fully connected linear layer with $\tanh$ activation functions. Similarly, the following hidden layers are rectified linear unit (ReLU) layers. The final two-layers is a vanilla linear layer followed by a normalization layer to satisfy the average power constraint. Recall that the input to the overall DNN-encoder are $m$-length subsequences of a binary codeword $c^{(j)}\in\Field_2^m$. The final normalization layer is applied as follows.  Define $\tilde \Cc_\mu$ as
\begin{align}
\tilde \Cc_\mu = \{ \tilde x: \tilde x = \tilde\mu_\theta(c^m),\, c^m\in\Field_2^m\},
\end{align}
where $\tilde \mu_\theta(c^m)$ is the output of the last linear layer, i.e., $\mu_\theta$ excluding the normalization layer. Let $\tilde \eta$ and $\tilde \sigma^2$ be the sample mean and variance of $\tilde \Cc_\mu$, respectively. Then, the normalization layer outputs are given by
\begin{align}
    x = \frac{\tilde x - \tilde \eta}{\tilde \sigma},
\end{align}
where $x$ is the output of the normalization layer.
Thus, the final normalization layer makes the constellation points satisfy the average power constraint. 
We note that the number of points in $\tilde \Cc_\mu$ is $2^m$ and the points are fixed once the parameters are fixed. In the training stage, the sample mean $\tilde \eta$ and variance $\tilde \sigma^2$ is updated for every parameter update (e.g. for each mini-batch stochastic gradient descent (SGD) update). After training, the normalization layer does not need any update since the parameters are then fixed.
Every layer except the final output layers of the encoder and decoder are implemented with batch normalization~\cite{pmlr-v37-ioffe15}.

\subsection{DNN inner-decoder (demodulator)}
The DNN-decoder (demodulator) has the following structure. First, the input of the DNN-decoder is formulated by a feature mapping function applied on the received signal $y$. Upon receiving the $j$-th channel output $y^{(j)}$, the received symbol is mapped to the logarithm of the channel distribution $p_{Y|X}$ as an $M$ dimensional vector, i.e., 
\begin{align}
    \psi(y^{(j)}, \sigma^2) = [\ln p_{Y|X}(y^{(j)}|x; \sigma^2): x\in \Cc_\mu]. \label{eq:psi}
\end{align} The DNN-decoder input $\psi(y^{(j)}, \sigma^2)$ is passed through some ReLU layers, and the final output layer is given by a sigmoid layer with $m$ output units representing the $m$ bit probabilities 
\begin{align}\label{eq:dnn-dec-output}
\phi_\theta(y^{(j)})= [p^{(j)}_1,\ldots, p_m^{(j)}],
\end{align}
 where $p_i^{(j)}\in[0,1]$ and we define $\phi_\theta(y^{(j)})$ as a shorthand notation for $\phi_\theta(\psi(y^{(j)}, \sigma^2))$.

\subsection{Loss function} \label{sec: loss}
For the loss function, we use an approximate variant of~\eqref{eq: gmi}. To this end, we first approximate~\eqref{eq: gmi} by
\begin{align}
\hat I^\text{gmi}(X;Y) &:= \E\left[\log\frac{q(X, Y)}{\sum_{x'\in \Cc_\mu }P_X(x')P_{Y|X}(Y|x')}\right] \label{eq:app1}\\
&= M+\E\left[\log\frac{q(X, Y)}{\sum_{x'\in \Cc_\mu }P_{Y|X}(Y|x')}\right]
\\
&= M+\E\left[\log\frac{\prod_{i=1}^m q(C_i, Y)}{\sum_{x'\in \Cc_\mu }P_{Y|X}(Y|x')}\right], \label{eq:loss_pre}
\end{align}
where we define the metric $q(x, y) = \prod_{i=1}^m q(c_i, y)$.
Note that we have defined the symbol level metric $q(x', Y)$ in the denominator of~\eqref{eq: gmi} by its optimal value $P_{Y|X}$ in~\eqref{eq:app1} and we further choose the symbol level decision metric $q(x, y)$ as a product of bit-level metrics $\prod_{i=1}^m q(c_i, y)$, which is a function of the output of the DNN-decoder. %The product form makes the metric compatible with a soft channel decoder. 
We note that the marginalization in the denominator of~\eqref{eq:loss_pre} is a function of the constellation points $\Cc_\mu$. Also, we treat the bit-level decision metric $q(c_i, y)$ to be estimates of $p(y|c_i)$ in product form to be compatible with linear code decoders, in particular, the sum-product algorithm.

In the following, we explain how we integrate the metric \eqref{eq:loss_pre} with our proposed architecture. Firstly, the channel input symbols are chosen as the outputs of the DNN encoder given by
\begin{align*}
x = \mu_\theta(c_1,\ldots, c_m),
\end{align*}
which results in $p_X(x) = 1/M$. For the decoding metric mapping, recall that our DNN decoder outputs are given by $\phi_\theta(y^{(j)})$ defined in~\eqref{eq:dnn-dec-output}. In our proposed architecture, the decision metric is defined by
\begin{align}
    q_\theta(c_i, y) = \begin{cases}
         p_i & \text{ if } c_i = 1\\
         1-p_i & \text{otherwise}.
    \end{cases}\label{eq:q}
\end{align}
We note that the metric $q_\theta(c_i, y)$ is a function of the DNN parameters $\theta$ since $p_i$ is the output of the overall DNN. When it is clear in the context, we will omit the subscript $\theta$ and use $q(c_i, y)$ for brevity.

The final step is to approximate $\hat I^{\text{gmi}}(X; Y)$ by its sample mean given as
\begin{align}
    \Lc(\theta) = M+\sum_{j=1}^N \frac{1}{N}\log\frac{\prod_{i=1}^m q_\theta(c_i^{(j)}, y^{(j)})}{\sum_{x'\in \Cc_\mu }P_{Y|X}(y^{(j)}|x')}, \label{eq:loss}
\end{align}
where $c_i^{(j)}$ is the $i$-th bit of the $j$-th input symbol, i.e., $x^{(j)} = \mu_\theta(c_1^{(j)}, \ldots, c_m^{(j)})$ and $y^{(j)}$ is the channel output of the $j$-th input symbol $x^{(j)}$. The DNN parameters $\theta$ are trained to maximize~\eqref{eq:loss}.

As a final note in this section, we point out some differences between the  binary cross-entropy (BCE) loss~\cite{Murphy2012} often used for binary classification and the loss given in equation~\eqref{eq:loss_pre}. In the Appendix, we show that the BCE loss is equal to a specialized GMI (up to a constant difference) when the metric is defined as an estimate of $p(c_i|y)$ while assuming that the bit probabilities are conditionally independent. We note that the conditional independence assumption is not true in general due to the linear encoding structure. Further improvements to account for the linear structure in the demodulation stage can be done by integrating iterative decoding and demodulation (IDD) as in~\cite{Cammerer--Aoudia--Dorner--Stark--Hoydis--Brink2020}. 
In our approach, we define our metric as an approximation of $p(y|c_i)$ in formulating the GMI which results in an additional term represented by the numerator inside the expectation of~\eqref{eq:loss_pre}. We note that the numerator term is a function of the DNN modulation constellation points which in turn makes it a function of the DNN parameters. Our proposed loss function explicitly utilizes the marginal $p(y)$ in the loss function and produces estimates of $p(y|c_i)$ that is directly compatible with SPA. Performance comparison on the BCE loss and our proposed metric are given in Section 4.

\subsection{Concatenation with linear codes}
Once the DNN is trained, we can combine it with outer linear codes. For ease of presentation, we will focus on LDPC codes, however, the structure can be combined with any binary linear code with efficient soft decoding algorithms. Let $c^n\in\Cc$ be a linear code codeword and assume $n=Nm$. The codewords are interleaved and demultiplexed into $m$-bits $c^{(j)} = (c^{(j)}_1,\ldots, c^{(j)}_m)$. The $j$-th subsequence is encoded by the DNN-encoder $\mu_\theta(c^{(j)})$, then sent through the channel to receive $y^{(j)}$, and finally decoded at the DNN-decoder $p^{(j)}=\phi_\theta(\psi(y^{(j)},\sigma^2) )$. The output of the decoder $p^{(j)}$ are then converted to decoding metrics $q(c_i^{(j)}, y^{(j)})$, $i=1,\ldots, m$~\eqref{eq:q} for training. For channel code decoding, the DNN outputs are translated to log-likelihood ratios (LLRs) by 
\begin{align}
    \ell^{(j)}_i = \ln \frac{1-p^{(j)}_i}{p^{(j)}_i},\,\, j=1,\ldots, N,\, i=1,\ldots, m. \label{eq:llr}
\end{align}
%\begin{align}
%    \ell^{(j)}_i = \ln \frac{1-q(c_i^{(k)}, y^{(k)})}{q(c_i^{(k)}, y^{(k)})},\,\, j=1,\ldots, N,\, i=1,\ldots, m. \label{eq:llr}
%\end{align}
Then, the LLRs are multiplexed into $n$-sequences and is decoded by the linear code decoder, e.g., sum-product algorithm. 

\section{Training methods and numerical evaluations}\label{sec: simulations}
The evaluations in this section have been implemented using the Pytorch~\cite{NEURIPS2019_9015} framework.
In our evaluations, we use the 5G standard LDPC codes~\cite{3gpp} with BG=1 and $Z_c=24$, rate $1/2$ LDPC codes which translate to length $1104$ (1056 after puncturing) bit codewords and use the sum-product decoding algorithm with 50 iterations for decoding of the outer code. 

Training of the DNN inner code is done independent of the linear channel code. That is, at training stage, $c^m$ are simply generated randomly and independently. For testing, we use the encoded bits from the linear code. We use a two-stage training process which we will refer to as the first stage training (pretraining) and the second stage training. 
The training parameters for each training steps are summarized in table~\ref{tab:dnn}. 

\begin{table}[!h]
\footnotesize
\centering
\caption{DNN training parameters for 1st stage and 2nd stage training.}
\begin{tabular}{@{}lcc@{}}
\toprule
Parameters & $M=16$ & $M=64$ \\ \midrule
Enc. hidden units                 & $[16, 64,  32]$        & $[64, 128, 128, 128]$        \\
1st stage dec. hidden units           & $[128]$      & $[128]$        \\
2nd stage dec. hidden units           & $[128]$         & $[64, 128]$         \\
Training $\SNR$           & $7$ dB      & $11.5$ dB        \\
1st stage batch size           & $M\times 20$      & $M\times 20$        \\
2nd stage batch size           & $M\times 1600$         & $M\times 1600$         \\
Number of samples           & $M\times 4800$      & $M\times 3200$        \\
Learning rate              & $0.1\sim 0.001$       & $0.1\sim 0.001$        \\
1st stage optimizer              & AdamW (0.01)       & AdamW (0.2)        \\
2nd stage optimizer              & Adam       & Adam        \\
Activation functions  & $\tanh$, ReLU    & $\tanh$, ReLU        \\ 
%Scheduling & cosine annealing & cosine annealing\\
\bottomrule
\end{tabular}
\label{tab:dnn}
\end{table}

\begin{figure}[!t]
\centering
\includegraphics[width=0.9\columnwidth]{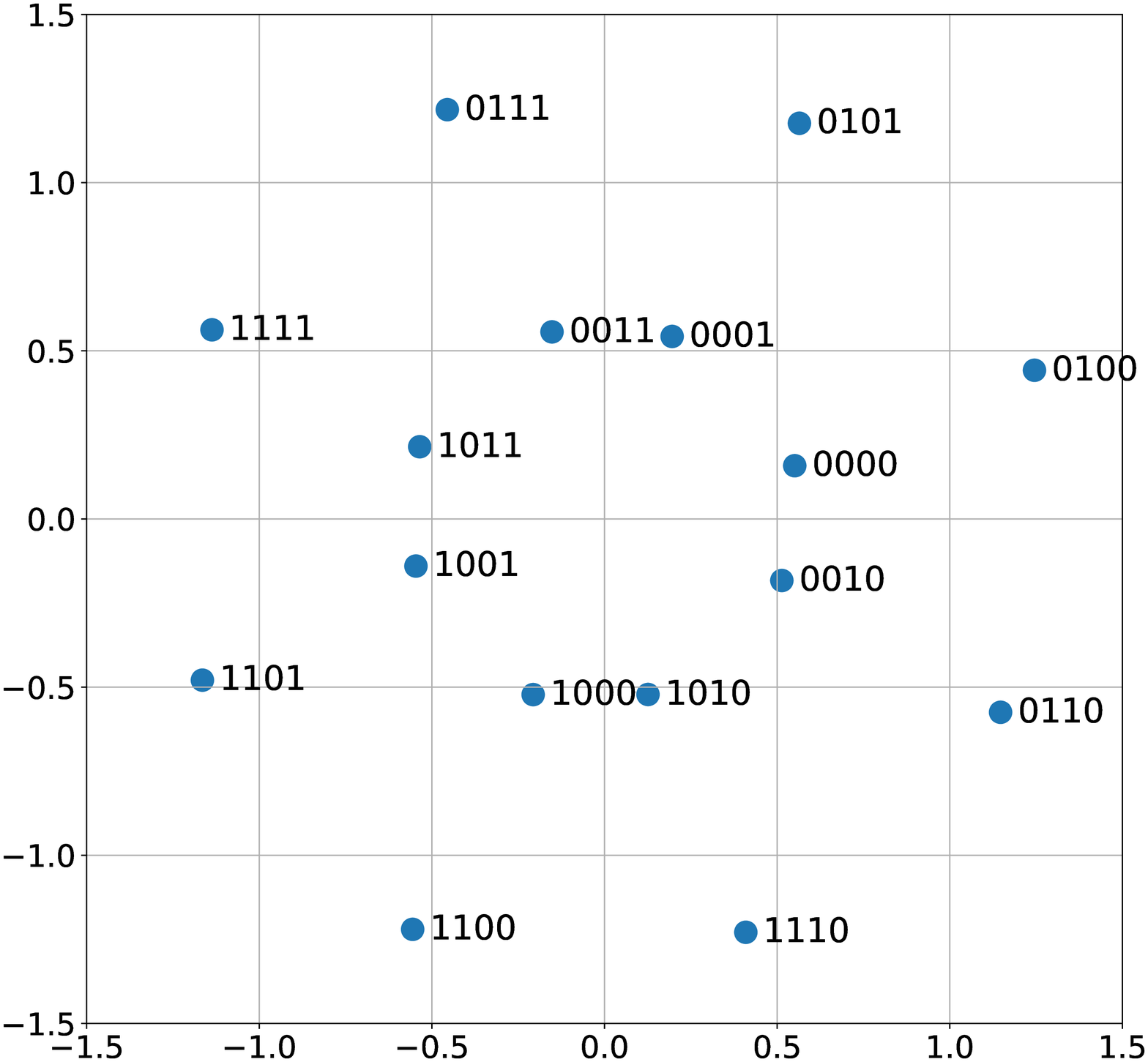}
\vspace{-2em}
\caption{Learned constellation $\Cc_\mu$ for $M=16$ with GMI maximization.}
\label{fig:constellation_16}
\end{figure}

\begin{figure}[!h]
\includegraphics[width=0.92\columnwidth]{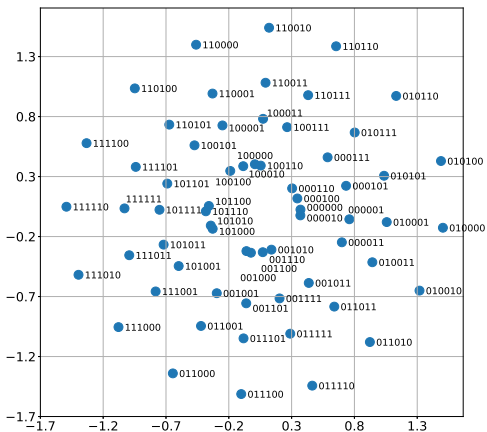}
\vspace{-2em}
\caption{Learned constellation $\Cc_\mu$ for $M=64$ with GMI maximization.}
\label{fig:constellation_64}
\vspace{-0.15em}
\end{figure}

In the first stage of training, we keep the decoding structure simple compared to the 2nd stage training, we have a relatively low batch size, and use the AdamW~\cite{loshchilov--hutter2018} optimizer with $L_2$ regularization with regularization coefficients given in the parenthesis of Table~\ref{tab:dnn}. These choice of hyper parameters help the optimizer escape local minima or saddle points to find a good initial constellation shape for further precise training at the second stage.

The second stage of training is used to fine-tune the DNN and train a better decoder via transfer learning. Firstly, we exchange the decoder with a randomly initialized larger set of decoding layers (while keeping the pretrained encoder). We train both the encoder and decoder parameters with a larger batch size and use the Adam optimizer without regularization. 

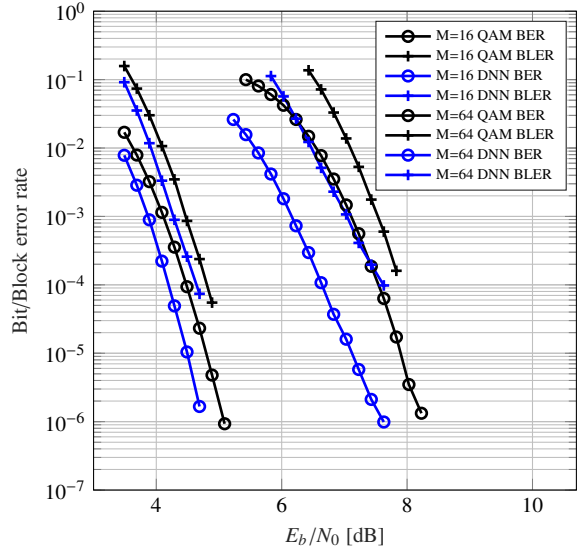
\begin{figure}[!t]
\vspace{-0em}
\centering
\begin{tikzpicture}
\footnotesize
\begin{axis}[%
width=2.5in,
height=2.5in,
at={(-1.0in,0.0in)},
scale only axis,
xmin=3,
xmax=10.7,
xlabel style={font=\color{white!15!black}},
xlabel={$E_b/N_0$ [dB]},
ymode=log,
ymin=1e-07,
ymax=1,
yminorticks=true,
ylabel style={font=\color{white!15!black}},
ylabel={Bit/Block error rate},
axis background/.style={fill=white},
xmajorgrids,
ymajorgrids,
yminorgrids,
legend style={nodes={scale=0.7, transform shape}, legend cell align=left, align=left, draw=white!12!black}
]
\addplot [color=black, line width=1.0pt, mark=o, mark options={solid, black}]
  table[row sep=crcr]{%
3.48970004336019	0.0169641435\\
3.68970004336019	0.00789192319\\
3.88970004336019	0.0032254073\\
4.08970004336019	0.0011455401\\
4.28970004336019	0.000355473496\\
4.48970004336019	9.39109921e-05\\
4.68970004336019	2.3221584e-05\\
4.88970004336019	4.79356153e-06\\
5.08970004336019	9.31818155e-07\\
};
\addlegendentry{M=16 QAM BER}

\addplot [color=black, line width=1.0pt, mark=+, mark options={solid, black}]
  table[row sep=crcr]{%
3.48970004336019	0.15787\\
3.68970004336019	0.07388501\\
3.88970004336019	0.03013\\
4.08970004336019	0.01067\\
4.28970004336019	0.00348\\
4.48970004336019	0.00086\\
4.68970004336019	0.000238000229\\
4.88970004336019	5.49999624e-05\\
};
\addlegendentry{M=16 QAM BLER}

\addplot [color=blue, line width=1.0pt, mark=o, mark options={solid, blue}]
  table[row sep=crcr]{%
3.48970004336019	0.00782726333\\
3.68970004336019	0.00285924226\\
3.88970004336019	0.000891884416\\
4.08970004336019	0.000222034063\\
4.28970004336019	4.92462143e-05\\
4.48970004336019	1.04356057e-05\\
4.68970004336019	1.6666665e-06\\
};
\addlegendentry{M=16 DNN BER}

\addplot [color=blue, line width=1.0pt, mark=+, mark options={solid, blue}]
  table[row sep=crcr]{%
3.48970004336019	0.0916099906\\
3.68970004336019	0.0353349984\\
3.88970004336019	0.011764998\\
4.08970004336019	0.00332999945\\
4.28970004336019	0.0008919999\\
4.48970004336019	0.000258000009\\
4.68970004336019	7.4000014e-05\\
};
\addlegendentry{M=16 DNN BLER}

\addplot [color=black, line width=1.0pt, mark=o, mark options={solid, black}]
  table[row sep=crcr]{%
5.42878723144531	0.0998038575053215\\
5.62878704071045	0.0807140097022057\\
5.8287878036499	0.0605499185621738\\
6.02878761291504	0.0421544425189495\\
6.22878742218018	0.0261450950056314\\
6.42878723144531	0.0148455146700144\\
6.62878704071045	0.00774661777541041\\
6.8287878036499	0.00352339027449489\\
7.02878761291504	0.00147247139830142\\
7.22878742218018	0.000560009328182787\\
7.42878723144531	0.000186407240107656\\
7.62878704071045	6.31959628663026e-05\\
7.8287878036499	1.72651471075369e-05\\
8.02878761291504	3.48958337781369e-06\\
8.22878742218018	1.32670447783312e-06\\
};
\addlegendentry{M=64 QAM BER}

\addplot [color=black, line width=1.0pt, mark=+, mark options={solid, black}]
  table[row sep=crcr]{%
6.42878723144531	0.136825010180473\\
6.62878704071045	0.0718750059604645\\
6.8287878036499	0.0330150052905083\\
7.02878761291504	0.013799998909235\\
7.22878742218018	0.00530500058084726\\
7.42878723144531	0.0017630037618801\\
7.62878704071045	0.000599498744122684\\
7.8287878036499	0.000160499839694239\\
};
\addlegendentry{M=64 QAM BLER}

\addplot [color=blue, line width=1.0pt, mark=o, mark options={solid, blue}]
  table[row sep=crcr]{%
5.22878742218018	0.0261261723935604\\
5.42878723144531	0.0156879648566246\\
5.62878704071045	0.00847771670669317\\
5.8287878036499	0.00416517024859786\\
6.02878761291504	0.00181583315134048\\
6.22878742218018	0.000730321975424886\\
6.42878723144531	0.000296136393444613\\
6.62878704071045	0.000107471612864174\\
6.8287878036499	3.70738634956069e-05\\
7.02878761291504	1.60416639118921e-05\\
7.22878742218018	5.78598474021419e-06\\
7.42878723144531	2.11363612834248e-06\\
7.62878704071045	9.88636315923941e-07\\
};
\addlegendentry{M=64 DNN BER}

\addplot [color=blue, line width=1.0pt, mark=+, mark options={solid, blue}]
  table[row sep=crcr]{%
5.8287878036499	0.112609997391701\\
6.02878761291504	0.0568100027740002\\
6.22878742218018	0.0268449988216162\\
6.42878723144531	0.0122549999505281\\
6.62878704071045	0.00512999948114157\\
6.8287878036499	0.00229999981820583\\
7.02878761291504	0.00106999999843538\\
7.22878742218018	0.000409999949624762\\
7.42878723144531	0.000196999913896434\\
7.62878704071045	9.79999822448008e-05\\
};
\addlegendentry{M=64 DNN BLER}

\end{axis}

% \begin{axis}[%
% width=5.174in,
% height=4.681in,
% at={(0in,0in)},
% scale only axis,
% xmin=0,
% xmax=1,
% ymin=0,
% ymax=1,
% axis line style={draw=none},
% ticks=none,
% axis x line*=bottom,
% axis y line*=left
% ]
% \end{axis}
\end{tikzpicture}%
\caption{Performance evaluations of block error and bit error rates for $M=16$ and $M=64$. The plots  compare our hybrid DNN strategy which maximizes GMI with a standard QAM coding scheme. The plots are results from the constellations given in Fig.~\ref{fig:constellation_16} and Fig.~\ref{fig:constellation_64}.}
\label{fig:ber_plots}
\end{figure}

In the following we present our numerical evaluations. In Fig.~\ref{fig:constellation_16} and Fig.~\ref{fig:constellation_64}, we show examples of trained constellation points $\Cc_\mu$ for $M=16$ and $M=64$, respectively. Notice that the constellation points have round edges compared to QAM constellations resulting in better shaping gains. It is interesting to note that the constellations seem to have structures with ``Gray mapping'' like labels, for example, the most significant bits (MSB) are distinguished by the left-half and the right-half of the plane.
In Fig.~\ref{fig:ber_plots}, we compare the bit-error rate (BER) and block-error rate (BLER) of our hybrid coded modulation strategy with the conventional QAM based BICM strategy. For hybrid coded modulation strategy provides approximately $0.3$ dB gain and $1$ dB gain over the conventional QAM based strategy for $M=16$ and $M=64$, respectively.

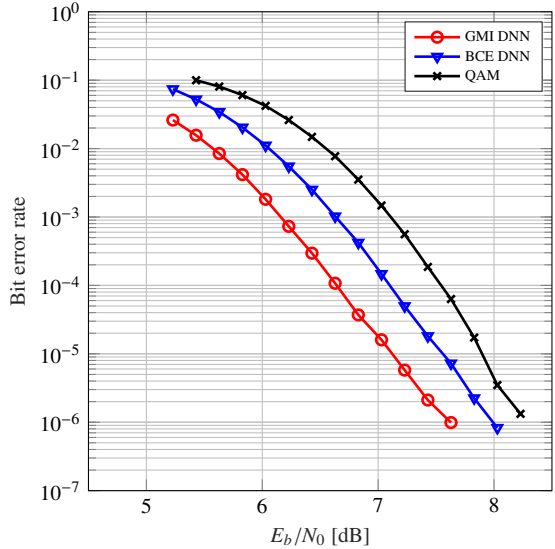
\begin{figure}[!h]
\begin{tikzpicture}
\footnotesize
\begin{axis}[%
width=2.4in,
height=2.5in,
at={(0.0in,0.0in)},
scale only axis,
xmin=4.5,
xmax=8.5,
xlabel style={font=\color{white!15!black}},
xlabel={$E_b/N_0$ [dB]},
ymode=log,
ymin=1e-07,
ymax=1,
yminorticks=true,
ylabel style={font=\color{white!15!black}},
ylabel={Bit error rate},
axis background/.style={fill=white},
xmajorgrids,
ymajorgrids,
yminorgrids,
legend style={nodes={scale=0.7, transform shape}, legend cell align=left, align=left, draw=white!12!black}
]
\addplot [color=red, line width=1.0pt, mark=o, mark options={solid, red}]
  table[row sep=crcr]{%
5.22878742218018	0.0261261723935604\\
5.42878723144531	0.0156879648566246\\
5.62878704071045	0.00847771670669317\\
5.8287878036499	0.00416517024859786\\
6.02878761291504	0.00181583315134048\\
6.22878742218018	0.000730321975424886\\
6.42878723144531	0.000296136393444613\\
6.62878704071045	0.000107471612864174\\
6.8287878036499	3.70738634956069e-05\\
7.02878761291504	1.60416639118921e-05\\
7.22878742218018	5.78598474021419e-06\\
7.42878723144531	2.11363612834248e-06\\
7.62878704071045	9.88636315923941e-07\\
};
\addlegendentry{GMI DNN}

\addplot [color=blue, line width=1.0pt, mark=triangle, mark options={solid, rotate=180, blue}]
  table[row sep=crcr]{%
5.22878745280338	0.0740861773\\
5.42878745280338	0.0524442554\\
5.62878745280337	0.0343669236\\
5.82878745280337	0.0203812033\\
6.02878745280337	0.0111144312\\
6.22878745280337	0.00550342873\\
6.42878745280337	0.00251505654\\
6.62878745280337	0.00102365538\\
6.82878745280337	0.000422073854\\
7.02878745280337	0.000147168571\\
7.22878745280337	4.98768874e-05\\
7.42878745280337	1.81572011e-05\\
7.62878745280337	7.24621117e-06\\
7.82878745280337	2.2462121e-06\\
8.02878745280337	8.27651384e-07\\
%8.22878745280337	4.71590865e-07\\
};
\addlegendentry{BCE DNN}

\addplot [color=black, line width=1.0pt, mark=x, mark options={solid, black}]
  table[row sep=crcr]{%
5.42878723144531	0.0998038575053215\\
5.62878704071045	0.0807140097022057\\
5.8287878036499	0.0605499185621738\\
6.02878761291504	0.0421544425189495\\
6.22878742218018	0.0261450950056314\\
6.42878723144531	0.0148455146700144\\
6.62878704071045	0.00774661777541041\\
6.8287878036499	0.00352339027449489\\
7.02878761291504	0.00147247139830142\\
7.22878742218018	0.000560009328182787\\
7.42878723144531	0.000186407240107656\\
7.62878704071045	6.31959628663026e-05\\
7.8287878036499	1.72651471075369e-05\\
8.02878761291504	3.48958337781369e-06\\
8.22878742218018	1.32670447783312e-06\\
};
\addlegendentry{QAM}

\end{axis}

\end{tikzpicture}% 
%\vspace{-2em}
\caption{Evaluation and comparison between GMI loss in~\eqref{eq:loss} and BCE for $M=64$.}
\label{fig:comparison}
\end{figure}

\section{Conclusion}\label{sec: conclusion}

In this paper, we propose a hybrid BICM architecture that combines binary linear codes with DNN based inner-codes. We formulate a GMI inspired loss function and design the architecture to be compatible with conventional linear codes and soft decoding algorithms. The inner DNN based code offers shaping gain compared to standard QAM based approaches while maintaining coding gains from practical length codes resulting in overall better error correcting performance.
Moreover, we provide some useful training methods for optimizing the DNN. Numerical results show that the proposed hybrid approach outperforms the QAM based BICM architectures which can provide gains for future high-order modulation communication systems. 

Some interesting future research directions would be to extend the framework to fading channels, multiple antennas, and multi-user channels.

\section*{Acknowledgments}\label{sec: ack}
This research was supported by the Hallym University Research Fund, 2019 (HRF-201910-012).
% This work was supported by the National Research Foundation of Korea (NRF) grant funded by the Korea government
% (MSIT) (No. 2020R1F1A107492612).

%% References


%\section{Acknowledgement}\label{sec: ack}

\section*{Appendix}\label{sec:app}
Consider the joint distribution $p_{C^m, X, Y}(c^m, x, y) = p(c^m)p(x|c^m)p(y|x)$ where $p(c^m)=1/2^{m}$, $c^m\in\Field_2^m$, $p(x|c^m)$ is the bit-to-symbol mapping distribution and $p(y|x)$ is the channel distribution. We define a decision metric in conditional distribution form as 
\begin{align}
    \tilde{q}(x|y) := \prod_{i=1}^m \tilde{q}(c_i|y). \label{eq:metric}
\end{align}
Then, specializing the GMI function with the metric \eqref{eq:metric}, we have
\begin{align*}
    I^\text{gmi}(X;Y) &= H(X) + \E\left[\log \tilde{q}(X|Y)\right]\\
    &= H(X) + \E\left[\log \prod_{i=1}^m \tilde{q}(C_i|Y)\right]\\
    &= H(X) + \sum_{i=1}^m \E\left[\log \tilde{q}(C_i|Y)\right]\\
    &= H(X) + \sum_{i=1}^m \E\left[\sum_{c_i}p(c_i|Y)\log \tilde{q}(c_i|Y)\right]\\
    &= H(X) - \sum_{i=1}^m\text{BCE}(p(c_i|Y), \tilde{q}(c_i|Y)),
\end{align*}
where the expectation is with respect to $p_{C^m, X, Y}$ and $\text{BCE}(p, q)$ is the binary cross entropy function. Thus, minimizing the BCE is equivalent to maximizing the GMI with decision metric in~\eqref{eq:metric}. 
%We note that the metric~\eqref{eq:metric} is not optimal in general.

\bibliographystyle{elsarticle-num}
% \bibliographystyle{elsarticle-harv}
% \bibliographystyle{elsarticle-num-names}
% \bibliographystyle{model1a-num-names}
% \bibliographystyle{model1b-num-names}
% \bibliographystyle{model1c-num-names}
% \bibliographystyle{model2-names}
% \bibliographystyle{model3a-num-names}
% \bibliographystyle{model3-num-names}
% \bibliographystyle{model4-names}
% \bibliographystyle{model5-names}
% \bibliographystyle{model6-num-names}

%\vspace{-0.3cm}

\bibliography{ict}
\end{document}